 \documentclass[preprint,floats,aps,epsfig,nofootinbib,amssymb]{revtex4}

\usepackage{slashed}
\usepackage{slashed}
\usepackage{graphicx,color}
\usepackage{epsfig}
\usepackage{subfigure}
\usepackage{epsfig}
\usepackage{dcolumn}
\usepackage{bm}
\usepackage{color}

\def\lsim{\mathrel{\rlap{\lower4pt\hbox{\hskip1pt$\sim$}}
    \raise1pt\hbox{$<$}}}         
\def\gsim{\mathrel{\rlap{\lower4pt\hbox{\hskip1pt$\sim$}}
    \raise1pt\hbox{$>$}}}         

\begin{document}

\vspace*{-5.8ex}
\hspace*{\fill}{NPAC-12-16}

\vspace*{+3.8ex}

\title{Higgs Vacuum Stability, Neutrino Mass, and Dark Matter}

\author{Wei Chao$^{1,2}$}
\email{chaow@physics.wisc.edu}
\author{Matthew Gonderinger$^4$}
\email{gonderinger@wayne.edu}
\author{Michael J. Ramsey-Musolf$^{1,3}$ }
\email{mjrm@physics.wisc.edu}
 \affiliation{$^1$ Department of Physics, University of Wisconsin-Madison, Madison, WI 53706, USA\\ 
 $^2$ Shanghai Jiao Tong University, Shanghai, China\\
$^3$California Institute of Technology, Pasadena, CA USA\\
$^4$Department of Physics and Astronomy, Wayne State University, Detroit, MI 48201  }

\vspace{3cm}

\begin{abstract}
Recent results from ATLAS and CMS point to a  narrow  range for the Higgs mass: $M_H\in[ 124,~126]~{\rm GeV}$.   Given  this range, a case may be made for new physics beyond the Standard Model (SM) because of the resultant vacuum stability problem, i.e., the SM Higgs quartic coupling may run to negative values at a scale below the Planck scale.  We study representative minimal extensions of the SM that can keep the SM Higgs vacuum stable to the Planck scale by introducing new scalar or fermion interactions at the TeV scale while solving other phenomenological problems. In particular, we consider the type-II seesaw model, which is introduced to explain the non-zero Majorana masses of the active neutrinos.  Similarly, we observe that if the stability of the SM Higgs vacuum is ensured by the running of the gauge sector couplings, then one may require a series of new electroweak multiplets, the neutral component of which can be cold dark matter candidate.  Stability may also point to  a new $U(1)$ gauge symmetry, in which the SM Higgs carries non-zero charge.

\end{abstract}

\maketitle
\section{Introduction}

In the Standard Model (SM) of particle physics,  the Higgs field \cite{higgs} provides the mechanism of spontaneous electroweak symmetry breaking and the origin of masses of the fundamental particles, but until recently the  Higgs boson itself left no signal in  high-energy collider experiments.  The situation has clearly changed, however, as both the ATLAS \cite{atlas} and CMS \cite{cms} collaborations have announced  observation of a bosonic particle at about the $5\sigma $ significance level.   The excess is driven by the   two channels with the highest mass resolution $H\to ZZ^{(*)} \to 4 \ell$ and $H\to \gamma \gamma$, and the equally sensitive but low resolution $H\to WW^{(*)}\to \ell \nu \ell \nu$ channel.  Assuming the boson spin is shown to be zero, these results will provide conclusive evidence for the discovery of a Higgs-like scalar particle with mass $126.0 \pm 0.4 ({\rm stat}) \pm 0.4 ({\rm sys}) ~{\rm GeV}$ for ATLAS and  $125.3 \pm 0.4 ({\rm stat}) \pm 0.5 ({\rm sys}) ~{\rm GeV}$ for CMS.  

The discovery points to a favored mass range: $M_H \in (124,~126)~{\rm GeV}$, which is in agreement with the indirect detections from the electroweak precision constraints,  $M_H < 158 ~{\rm GeV}$ \cite{ewpm}. It  fixes the one remaining free parameter in the SM:  the Higgs self-coupling $\lambda$.\footnote{The Higgs quadratic coupling can be expressed in terms of the Higgs vacuum expectation value and $\lambda$.}   However, this low Higgs mass immediately leads to the problem of the SM Higgs vacuum stability which  requires $\lambda$ remain positive at all scales $\Lambda$. If $\lambda$ becomes negative at some scale, the potential is either unbounded from below and has no state of minimum energy or has a vacuum with lower energy for the case where $\lambda$ may run positive again at an even higher scale. Given $M_H \sim 126 ~{\rm GeV}$, the Higgs self-coupling $\lambda$ may run negative at a scale below the Planck scale~\cite{Holthausen:2011aa,strumia,xing2}, necessitating new physics beyond the SM (BSM). It was shown in \cite{Bezrukov:2012sa} that  absolute vacuum stability requires a Higgs mass $M_H\ge 129\pm 6~{\rm GeV}$, by using a partial two-loop matching and three-loop renormalization group running procedure and taking into account the existing $2\sigma$ experimental uncertainties in the mass of the top quark and $\alpha_s$.  A very similar conclusion was given in Ref. \cite{Alekhin:2012py}.  Additionally, Ref.~\cite{Degrassi:2012ry} studied the two-loop QCD and Yukawa corrections to the relation between the Higgs quartic coupling and the Higgs mass so as to reduce the uncertainty in the determination of the Higgs mass from $\lambda(\mu)$.  The authors claimed that while $\lambda$ at the Planck scale is zero, the absolute stability of the SM Higgs potential is excluded at $98\%$ C.L. for $M_H<126~{\rm GeV}$. Thus a $\sim$124-126~GeV Higgs strongly points to new physics in the desert between the Fermi and Planck scales. Supersymmetry and extra dimension models may provide solutions to this problem, but up to now no clear signals of such new physics have emerged from collider searches.  Therefore it is instructive to consider alternative simple extensions of the SM that can alleviate the vacuum stability problem.

Despite the absence of collider BSM signals, definitive evidence of new physics beyond the SM comes from several additional sources,  including neutrino masses and dark matter.  The solar, atmospheric, reactor, and accelerator neutrino experiments have provided  convincing evidence that neutrinos are massive and lepton flavors are mixed \cite{pdg}.   Precise cosmological observations have confirmed the existence of non-baryonic cold dark matter with an abundance of $\Omega_D h^2 = 0.1123\pm0.0035$ \cite{wmap}. These two important discoveries cannot be accommodated in the SM without introducing extra ingredients. Doing so can affect the stability of the electroweak vacuum {\em via} one or more of the following interactions: (1) additional quartic scalar interactions associated with new scalar degrees of freedom; (2) Yukawa interactions associated with neutrino mass generation; (3) modified or extended gauge interactions appearing in neutrino mass and/or dark matter models.

In this work, we study examples of all three. In particular, to understand the origin of the neutrino masses, one may extend the SM with heavy Majorana neutrinos  so that light neutrino masses can  be generated through  the so-called seesaw mechanism.  There are three types of tree-level seesaw mechanisms, categorized according to the particle content of their extension to the SM: heavy Majorana neutrinos (Type-I \cite{seesaw1}) plus either a $Y=1$ Higgs triplet (Type-II \cite{seesaw2}) or a $Y=0$ Fermion triplet (Type-III \cite{seesaw3})  ($Y$ is the SM hypercharge quantum number).  In the context of Type-I models, the impact of heavy right-handed neutrinos, $N_R^{}$, on the SM Higgs vacuum stability and metastability scales was studied in Ref.~\cite{strumia,Rodejohann2012,Masina:2012tz}, with the result that the $N_R^{} $ decrease the Higgs vacuum stability scale.  In this paper we will study the effect of a TeV scale Type-II seesaw model on the Higgs vacuum stability. Our result shows that the the SM Higgs vacuum can remain stable up to the Planck scale for certain chosen parameters of the Type-II seesaw model.

We also study how to keep the SM Higgs vacuum stable by making changes only to the gauge interactions. In the context of the SM gauge groups, stability  up to the Planck scale can be restored by introducing a series of new electroweak multiplets, the neutral component of which may serve as the cold dark matter candidate.
The presence of these  multiplets may significantly alter the renormalization group (RG) evolution of the gauge coupling coefficients and thus indirectly keep the SM Higgs vacuum stable up to the Planck scale.   We also study the implications of the SM Higgs vacuum stability for a new gauge symmetry.  In Ref.~\cite{hexg} it was shown that there are only flavor-dependent anomaly-free gauged $U(1)$ symmetries in the minimal SM. In the SM with three right-handed neutrinos and a scalar singlet,  there is another well-known anomaly-free gauged $U(1)$ symmetry: $U(1)_{B-L}$ \cite{bminusl}.  However, the SM Higgs boson carries no new $U(1)$ charge in both cases.   The presence of these new $U(1)$ symmetries  only affects the running behavior of the Yukawa couplings (most notably that of the top quark).  In this paper, we study the effect of a new type of $U(1)$  gauge symmetry in which  the SM Higgs boson carries non-zero charge.  The anomalies are spontaneously cancelled by extending the SM with three right-handed neutrinos.   We derive the  one-loop $\beta$-functions of the model and investigate their effects on the SM Higgs vacuum stability. We find that stability of the vacuum can be achieved for some regions of parameter space.

The paper is organized as follows: in section II we show preliminary formulas relevant for the numerical analysis.  We investigate the Type-II seesaw effect in section III.  Section IV is devoted to studying the implications of the SM Higgs vacuum stability for the gauge interactions.  We summarize in section V. Expressions for the relevant $\beta$-functions appear in the Appendix.

\section{ Preliminaries }

We first review the stability analysis within the SM.  A constraint on the Higgs mass can be obtained by the requirement that spontaneous symmetry breaking actually occurs, that is,  $V(v)$ be the minimum of the Higgs potential
\begin{eqnarray}
V(H)=-\mu^2 H^\dagger H + {1 \over 2} \lambda (H^\dagger H)^2 \ \ \ ,
\end{eqnarray}
where  $v\approx 246$ GeV is the Higgs vacuum expectation value (VEV).  This bound is essentially equivalent to the requirement that the quartic Higgs coupling  $\lambda(\mu)$ never becomes negative at  any scale $\mu < \Lambda_{\rm NP}$ , where $\Lambda_{\rm NP}$ is the scale of new physics.\footnote{For the first paper to include precise evaluation of the renormalization group evolution when studying the SM Higgs vacuum stability problem, see \cite{Lindner:1988ww}.}   In this paper we will study numerically the impact of representative simple BSM scenarios on
 the vacuum stability of a $\sim$124-126 GeV SM Higgs. As a SM baseline, we will use  the two-loop beta functions of the Higgs self coupling $\lambda$, the gauge couplings $g_i$ $(i=1,2,3)$ and the top quark Yukawa coupling $y_t$, as well as the one-loop matching condition for the SM  Higgs mass. Contributions of BSM physics are considered at the one-loop level. The resulting stability requirements for a given BSM scenario are likely to be overly conservative, since the three-loop analyses tend to alleviate the tension of a 124-126 GeV Higgs with stability. However, since we do not presently have in hand the two-loop running for the BSM scenarios, it may not make sense to consider the SM at the three-loop and the BSM at the one-loop. Thus, for illustrative purpose we will use the ``one-loop matching and two-loop renormalization group running" procedure. The $\beta$-function of $\lambda $ is given to two loop order in Eq. (\ref{lambdabeta}), in which $\beta_\lambda^{(1)}, \beta_\lambda^{(2)}$ represent  the $\beta$-functions of $\lambda$ at the one-loop and two-loop level, $t\equiv \ln \mu/\mu_0$ with $\mu_0$ being a reference energy scale, and $y_t$ is the Yukawa coupling of the top quark.  For illustrative purposes, we neglect the scale-dependence of the Yukawa and gauge couplings, arriving at a simplified condition for vacuum stability at the scale $\mu$:
\begin{eqnarray}
\lambda(\mu) \approx \lambda (\Lambda_{\rm EW}) +  \left(  {1 \over 16 \pi^2} \beta_\lambda^{(1)} + { 1 \over (16 \pi^2)^2} \beta_\lambda^{(2)} \right ) \ln \left ( {\mu \over \Lambda_{\rm EW} }\right)  > 0 \; .
\end{eqnarray}
A meaningful and complete analysis should take into account the running behavior of all parameters.  One should also take implement the one-loop matching condition between the running Higgs quartic coupling and the Higgs boson pole mass $M_H$ \cite{hambye}: 
 \begin{eqnarray}
 \lambda(M_H) v^2 = M_H^2 \left[ 1+ \Delta (M_H^{} ) \right] \; .
 \end{eqnarray}
where the expression of $\Delta(x)$ can be found in Ref. \cite{sirlin}.

The  two loop $\beta$-functions for the gauge couplings are given in Eq. (\ref{betag}), in which $Y_{U,D,E}$ represent the Yukawa coupling matrices of up-quarks, down-quarks, and charged leptons, respectively. Here the $SU(3)_C \times SU(2)_L \times U(1)_Y$ gauge couplings $g_3, ~g_2, ~g_1$ are normalized based on $SU(5)$ (though we do not impose any GUT relations on the couplings), so the electroweak couplings $g$ and $g^\prime$ are related to these by $g^2 =g_2^2 $ and ${g^\prime}^2 = (3/5) g_1^2 $. The determination of the  couplings proceeds from the relations $\alpha_i  \equiv g_i^2 /4\pi$, with $(\alpha_1, ~\alpha_2, ~ \alpha_3) = (0.01681,~0.03354,~0.1176)$ at the Z-pole \cite{pdg}.

The two-loop $\beta$-function of the top quark Yukawa coupling  is given in Eq. (\ref{betatop}). The initial input of $y_t$ is given by $y_t (M_t) = \sqrt{2} m_t(M_t) /v$,  where $v=246.2 ~{\rm GeV}$ and  $m_t$ is the  top quark running mass determined from~\cite{beta}:
\begin{eqnarray}
M_t \approx m_t (M_t) \left( 1 + {4 \over 3} {\alpha_3 (M_t ) \over \pi} + 11 \left(  {\alpha_3 (M_t) \over \pi }\right)^2 - \left(  {m_t(M_t) \over 2 \pi v }\right) ^2
 \right)  \; , \label{match}
 \end{eqnarray}
in which the second and the third terms correspond to the one- and two-loop QCD corrections while the fourth term comes from the electroweak corrections at the one-loop level.   We use the running mass of the top quark value $m_t(M_Z)=172.1$ GeV \cite{xing2} in our following numerical analysis.  Utilizing the foregoing RG analysis and the present range for $M_H$, one finds that $\lambda$ runs negative for $\Lambda\sim 10^9 - 10^{11}$ GeV.

\section{ a new scalar interaction and neutrino mass}

A simple solution to the 125 GeV SM Higgs vacuum stability problem is obtained by introducing a new beyond-the-SM scalar that may interact with the SM Higgs through a four scalar coupling vertex. 
Typical examples are the Higgs portal dark matter models, e.g. the scalar singlet or  \lq\lq darkon"~\cite{Barger:2007im,Barger:2008jx,Cai:2011kb} and inert dark matter models~\cite{inertdm}. For a detailed analyses of the implications of these models on the  Higgs vacuum stability,  see Refs.~\cite{raidal,taiwan,higgsportal,threshold,gondering,Chun:2012jw, Cheung:2012nb}. 

The fact that neutrinos have tiny but non-zero masses is the first  (terrestrial) experimental evidence of new physics beyond the SM.  The most convincing idea to understand the origin of neutrino masses is the seesaw mechanism.  The effect of a Type-I seesaw model on the Higgs vacuum stability was studied in Ref. \cite{strumia,Rodejohann2012}, and the Type-I seesaw model was found to aggravate the instability of the vacuum.  Here we investigate the effect of a TeV scale Type-II seesaw model, which extends the SM with a  triplet scalar $\Delta$, 
\begin{eqnarray}
\Delta = \left( \matrix{\Delta^{+}/\sqrt{2} & \Delta^{++} \cr \Delta^0 & - \Delta^+ /\sqrt{2} } \right) \; , \nonumber
\end{eqnarray}
transforming as (3,1) under the electroweak gauge group $SU(2)_L \times U(1)_Y$.  The additional scalar potential can be written as
\begin{eqnarray}
\Delta V =&& M_\Delta^2 {\rm Tr } [\Delta^\dagger \Delta] + {\lambda_1 \over 2 } \left(  {\rm Tr} [\Delta^\dagger \Delta ]\right)^2 + {\lambda_2 \over 2 } \left[  ({\rm Tr } [\Delta^\dagger \Delta])^2 -{\rm Tr} (\Delta^\dagger \Delta \Delta^\dagger \Delta)\right] \nonumber \\
&+& \lambda_4 H^\dagger H {\rm Tr } [\Delta^\dagger \Delta] + \lambda_5 H^\dagger [\Delta^\dagger,~ \Delta] H + \left[ \sqrt{2}\lambda_6  H^T i \sigma_2 \Delta^\dagger H + {\rm h.c.} \right] \; .
\end{eqnarray}
The scalar triplet couples to the left-handed lepton doublet through the following Yukawa interaction:
\begin{eqnarray}
-{1\over \sqrt{2} } \left(Y_\Delta\right)_{ij} \overline{\ell_{L}^{C i}} \varepsilon \Delta \ell_L^j + {\rm h.c.} \; , \label{neumass}
\end{eqnarray}
where $C$ is the charge conjugation operator. The active neutrino mass can be derived from Eq. (\ref{neumass}) after the spontaneous breaking of electroweak symmetry: $ M_\nu = Y_\Delta v_\Delta$ where $v_\Delta$ is the VEV of the scalar triplet and is constrained to be less than $1~{\rm GeV}$ by the $\rho$ parameter.

 In this model  the one-loop  $\beta$-function of $\lambda$ can be written as \cite{chao,schmidt}
\begin{eqnarray}\label{eq:beta_lambda_seesaw}
\beta_\lambda^{(1)} = \left( \beta_\lambda^{(1)}\right)_{\rm SM} + 6 \lambda_4^2 + 4 \lambda_5^2 \; ,
\label{belam}
\end{eqnarray}
and the one loop $\beta$-functions for the gauge couplings in this case are given by $(b_1^{}, ~b_2^{}, ~b_3^{})=(47/10,~-5/2,~-7)$, with $b_i$  defined in Eq. (\ref{betag}). Notice that  both $\lambda_4$ and $\lambda_5$ contribute to the $\beta$-function of $\lambda$.  Here we mainly consider the effect of $\lambda_4$ by working in the limit where the other triplet couplings ($\lambda_1, \lambda_2$, and $\lambda_5$) are equal to zero at the input scale.  This is a reasonable simplification since $\lambda_1$ and $\lambda_2$ do not contribute to $\beta_\lambda$ at one-loop level, and as Eq. (\ref{eq:beta_lambda_seesaw}) shows $\lambda_4$ has a larger impact on the running of $\lambda$ than $\lambda_5$ (assuming $\lambda_4$ and $\lambda_5$ are of the same order).\footnote{ Equivalent conclusions can be reached by considering the effect of $\lambda_5$ as well.}  Nonetheless we do include the RG evolution of all the scalar couplings in our analysis and we ensure that all of the vacuum stability conditions in Ref. \cite{Arhrib:2011uy}, which in our notation are given by
\begin{eqnarray}
&\lambda>0\label{eq:seesaw_stability_1}\\ 
&\lambda_1\geq 0\label{eq:seesaw_stability_2}\\ 
&\lambda_1 + {1\over  2} \lambda_2 >0\label{eq:seesaw_stability_3}\\ 
&\lambda_4 (+ \lambda_5) + {1\over 2}\sqrt{\lambda \lambda_1} >0 \label{eq:seesaw_stability_4}\\ 
&\lambda_4 (+ \lambda_5) +{1\over 2} \sqrt{\lambda (\lambda_1 + {1\over 2 } \lambda_2)} >0 \label{eq:seesaw_stability_5}\ \ ,
\end{eqnarray}
are satisfied for all values of $\mu$ between $M_H$ and $M_{pl}$ (the parentheses in Eqs. (\ref{eq:seesaw_stability_4}) and (\ref{eq:seesaw_stability_5}) indicate that there are actually two stability conditions in each equation: one with $\lambda_5$ in the parentheses taken into account and the other one without).  A vacuum stability analysis in which the other Type-II seesaw scalar couplings are allowed to have non-zero values at the input scale is much more complicated (see \cite{Chun:2012jw}) and would distract from our purpose of studying the RG evolution of $\lambda$.  Furthermore, we note that, as $\lambda_6$ appears in neither the other scalar coupling $\beta$-functions nor the vacuum stability conditions in Eqs.~(\ref{eq:seesaw_stability_1})-(\ref{eq:seesaw_stability_5}), we do not include it in our analysis.  The coupling $\lambda_6$ effects the seesaw mechanism by giving the triplet a VEV:
\begin{equation}
v_\Delta \sim \frac{\lambda_6 v^2}{ M_\Delta^2}
\end{equation} and as mentioned above $M_\nu = Y_\Delta v_\Delta$.  Bounds on $\lambda_6$ that arise from avoiding tachyonic directions in the potential at the EW minimum \cite{Arhrib:2011uy} can always be satisfied -- while still obtaining the desired neutrino masses -- through appropriate choices of the triplet mass scale $M_\Delta$ and the Yukawa couplings $Y_\Delta$.

\begin{figure}[h!]
\includegraphics[width=8cm]{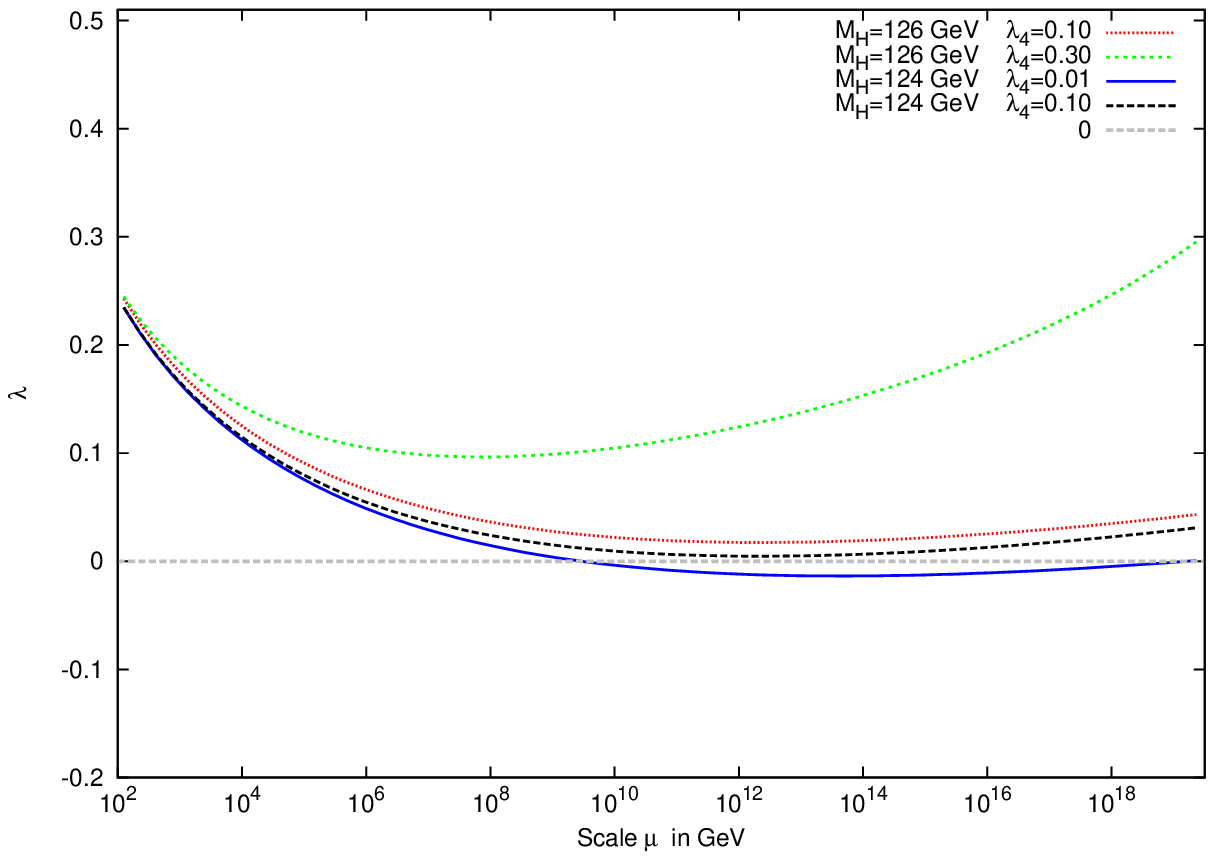}
\includegraphics[width=8cm]{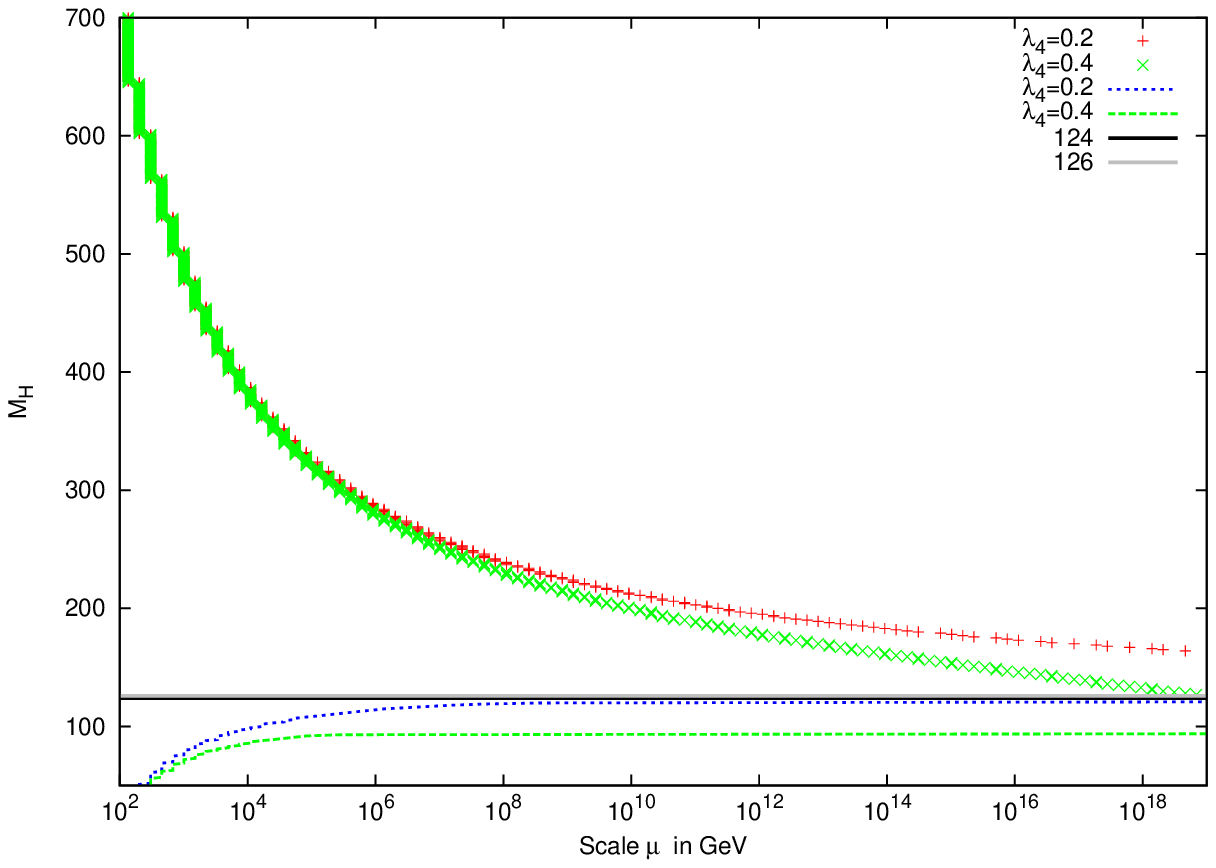}
\caption{Left panel: $\lambda$ as the function of energy scale $\mu$. The solid and dashed correspond to $M_H=124~{\rm GeV}$ and $\lambda_4=0.01, 0.10$ respectively.  The short-dashed and dotted lines correspond to $M_H=126~{\rm GeV}$ and $\lambda_4=0.30, 0.10$ respectively. 
Right panel: $M_H$ as the function of cutoff scale $\mu$. The plus and cross signs are the perturbativity constraint with $\lambda_4$ being $0.2$ and $0.4$ . The dashed and dotted lines are the vacuum stability constraint with $\lambda_4=0.4$ and $0.2$,  respectively.  The horizontal  band is the current experimental value of Higgs boson mass.}
\label{triplet}
\end{figure}

To study the Higgs vacuum stability, we first calculate the quartic coupling $\lambda (M_H)$ using the one-loop matching condition in Eq. (\ref{match}), then run $\lambda(\mu)$ to the Planck scale by solving the RG equations. The $\beta$-function of $\lambda_4$ is 
\begin{eqnarray}
16 \pi \beta_{\lambda_4} =&& -\left({9\over 2} g_1^2 +{33 \over 2} g_2^2 \right)\lambda_4+ 6 g_2^4 + {27 \over 25 }g_1^4 + 8 \lambda_5^2 \nonumber \\
&& + (8\lambda_1 +2 \lambda_2 + 6 \lambda +4\lambda_4 +2{\rm Tr}[Y_\Delta^\dagger Y_\Delta] +6 y_t^2) \lambda_4 \; . \label{phi}
\end{eqnarray}
The $\beta$-function for the top Yukawa coupling is  the same as that in the SM.  The $\beta$-functions for the $\lambda_i$ $(i=1,2,5)$ can be found in the appendix.  We assume the scalar triplet is at the electroweak scale and therefore $Y_\Delta \ll 1$, as implied by the scale of the light neutrino masses. Consequently we can safely neglect the ${\rm Tr} [Y_\Delta^\dagger Y_\Delta]$ term in Eq. (\ref{phi}) and need not  consider the matching condition at the seesaw threshold.  In the left panel of FIG. \ref{triplet} we plot $\lambda$ as a function of energy scale $\mu$. The solid and dashed lines correspond to $M_H=124~{\rm GeV}$ and $\lambda_4=0.01, 0.10$, respectively.  The short-dashed and dotted lines correspond to $M_H=126~{\rm GeV}$ and $\lambda_4=0.30, 0.10$. 
We find that the vacuum of a 124-126~GeV Higgs can be stable up to the Planck scale for the case of $\lambda_4=0.1$, while the vacuum of the  $124~{\rm GeV}$ Higgs will be unstable at the scale of ${\cal  O}(10^{10}) ~{\rm GeV}$ for the case $\lambda_4=0.01$. 

It is  interesting and instructive to also study the perturbativity constraints in this model. The perturbativity bound is defined as the highest Higgs boson running mass given by the the Higgs quartic coupling which satisfies the condition $\lambda(\mu)< 8.2$ \cite{perturbativity} for any $\mu$ between the electroweak and Planck scale, $M_{pl}$ (this criterion is less stringent than that used in \cite{gondering} which was based on the work of \cite{Riesselmann:1996is}).  With this perturbativity requirement, in conjunction with the vacuum stability conditions, we plot in the right panel of FIG. \ref{triplet} the bounds on the Higgs mass $M_H$ as a function of the energy scale $\mu$. The plus and cross signs are the perturbativity constraints with $\lambda_4$ equal to $0.2$ and $0.4$, respectively. The dashed and dotted lines are the vacuum stability constraints with $\lambda_4=0.4$ and $0.2$,  respectively.  The horizontal  band is the current experimental value of Higgs boson mass. We can read from the figure that the 125~GeV SM Higgs mass satisfies both the perturbativity and the vacuum stability constraints up to the Planck scale for appropriately chosen initial input value of $\lambda_4$.  Note that the range of this coupling that is consistent with both vacuum stability and perturbativity is rather restricted.

We comment that our analysis is similar in spirit to that of Ref.~\cite{Chun:2012jw}, though with some differences. In particular, we use  two-loop $\beta$-functions for the gauge and Yukawa couplings, analyze in more detail the perturbativity bounds, and a study of a range of input values for $\lambda_4$. On the other hand, \cite{Chun:2012jw} considers the behavior of the full set of scalar couplings, constraints for electroweak precision data, and implications for the $H\to\gamma\gamma$ rate.

\section{Modified Gauge Interactions }

In this section, we consider alternative solutions to the SM Higgs vacuum stability problem by (a) modifying the  $\beta$-functions of the gauge couplings  through the introduction of new electroweak (EW) multiplets, which might provide a cold dark matter candidate,  or (b) introducing a new  $U(1)$ gauge symmetry.

\subsection{Higgs vacuum stability with new EW multiplet}
Higher representation  EW multiplets exist in various models. A typical example is the gauge portal dark matter model in which dark matter annihilates into the SM particles through EW gauge interactions.  The gauge portal scenario is one genre of a more general set of dark matter models that also include Higgs portal and axion portal models among others.   Minimal dark matter \cite{minimal, yicai,Kumericki:2012bf} is a typical gauge-portal dark matter model in which a high-dimension electroweak multiplet ({\em e.g.}, a $(1,~5,~0)$) with hypercharge $Y=0$ is introduced.  There are also models where an electroweak  triplet \cite{FileviezPerez:2008bj,chaotri} or 7-plet \cite{septuplet} can be dark matter candidates.\footnote{Recently,  it was observed that the viability of scalar dark matter in these scenarios typically requires the introduction of an additional discrete symmetry in order to avoid the presence of destabilizing super-renormalizable interactions\cite{Kumericki:2012bf}.}   Taking into account loop contributions from these new EW multiplets, the one-loop $\beta$-functions of $g_2$ and $g_1$  will be
\begin{eqnarray}
&&16 \pi^2 \beta_{g_2}^{(1)} =16 \pi^2 \left(\beta_{g_2}^{(1)}\right)_{\rm SM} +{n (n^2-1)\over 18(1+ \zeta) } g_2^3  \; , \\
&& 16 \pi^2 \beta_{g_1^{}} ^{(1)}= 16 \pi^2\left(\beta_{g_1}^{(1)}\right)_{\rm SM} + {2n \over 5(1+\zeta)  } Y^2 g_1^3 \; .
\end{eqnarray}
where $\zeta=1 ~{\rm or}~0$ for bosonic or fermionic dark matter respectively. $Y$ is the weak hypercharge of the dark matter and $n$ is the dimension of the $SU(2)_L$ multiplet representation.
We can conclude that the running behavior of the $g_i$ may be significantly changed for these  cases and thus may have an effect on the RG evolution of the Higgs quartic coupling.

\begin{figure}[h!]
\includegraphics[width=8cm]{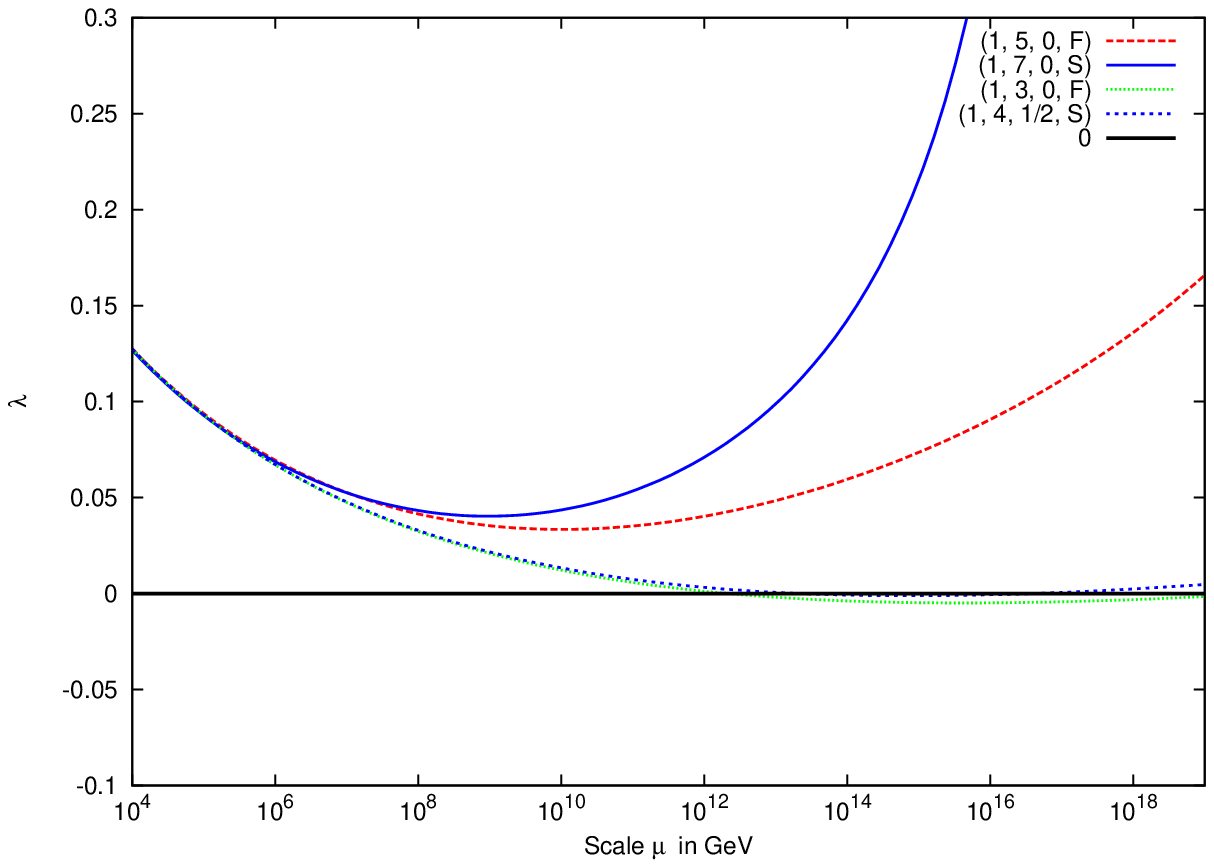}
\includegraphics[width=8cm]{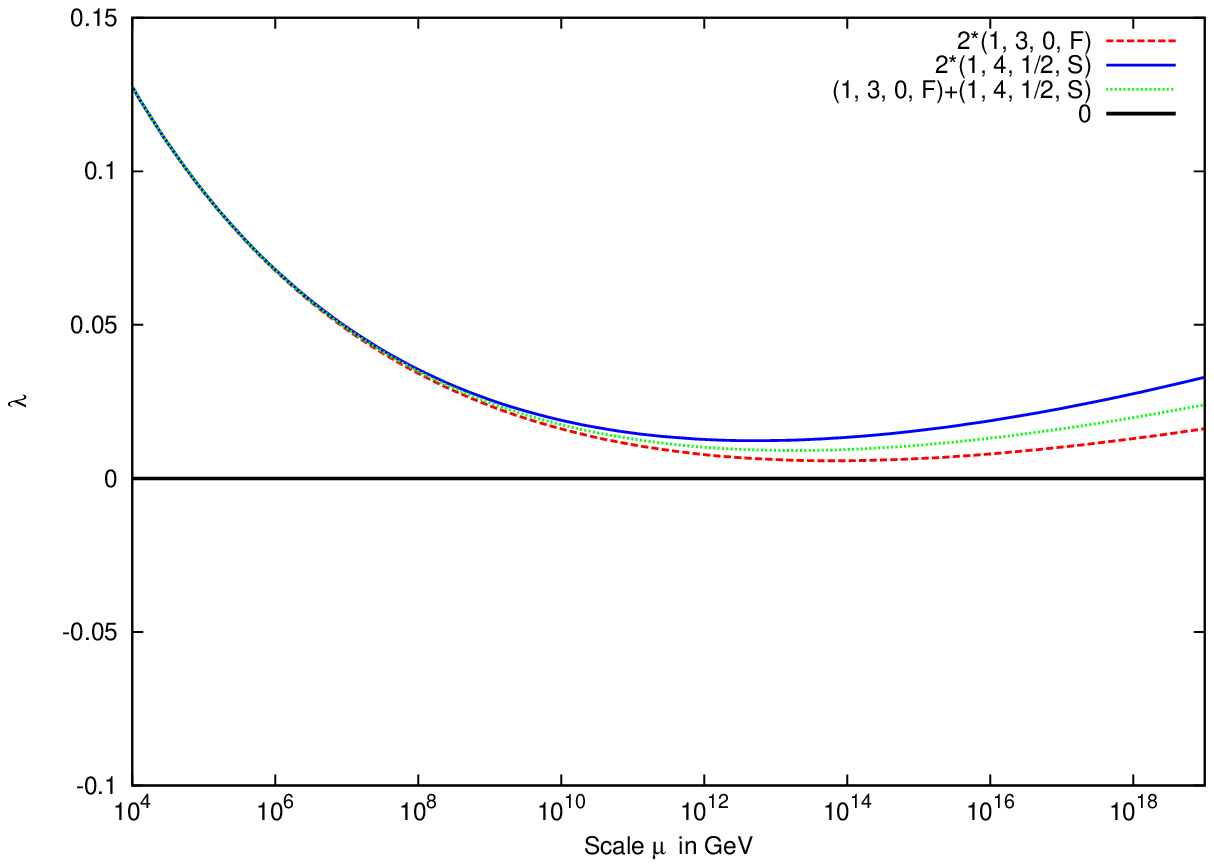}
\caption{Left panel: $\lambda$ as the function of energy scale $\mu$, where the solid, dashed, dotted and shot-dashed lines correspond to $(1,~7,~0, ~S )$, $(1,~5,~0, ~F )$, $(1,~4,~1/2, ~S )$ and $(1,~3,~0, ~F )$ EW multiplet cases respectively, where $S$ and $F$ represent bosonic or fermionic fields. Right panel: $\lambda$ as the function of $\mu$, the solid, dashed and dotted lines correspond to two quadruplet, two triplet and one triplet plus one quadruplet cases, respectively. }
\label{multiplet}
\end{figure}

To illustrate, we plot in the left-panel of FIG. \ref{multiplet} the coupling $\lambda$ as the function of energy scale $\mu$, where the solid, dashed, dotted,  short-dotted and dashed lines correspond to the addition of $(1,~7,~0, ~S )$, $(1,~5,~0, ~F )$, $(1,~4,~1/2, ~S )$ and $(1,~3,~0, ~F )$  EW multiplets, separately ($S$ and $F$ represent bosonic --- i.e., scalar --- and fermionic fields). We set $M_H=126~{\rm GeV}$  and $m_t (M_H) =167~{\rm GeV}$ for the initial input.  The mass of the EW multiplet is set to be several TeV so as to be consistent with the constraint on the dark matter relic abundance \cite{minimal,chaotri}.  We conclude from the figure that the $126 ~{\rm GeV}$ SM Higgs vacuum is stable up to the Planck scale with the addition of a single 5-plet or 7-plet; however, the Higgs vacuum remains unstable with the addition of a single electroweak triplet or quadruplet.  At least two extra EW multiplets of the triplet or quadruplet variety are needed to keep the Higgs vacuum stable up to the Planck scale as can be seen from the right panel of  FIG. \ref{multiplet}. 

\subsection{ Higgs vacuum stability with Extra $U(1)$ gauge symmetry} 

It has been shown that there are only flavor dependent anomaly-free gauged $U(1)$ symmetries, i.e., $U(1)_{L_i-L_j}$ \cite{hexg},  in the SM.  A $U(1)_{B-L}$ gauge symmetry emerges in the SM plus three right-handed neutrinos.\footnote{The additional $U(1)$ may also be global, but  we focus on the gauged case.}   However, the SM Higgs boson carries no $U(1)$ charge for both cases,  so the only impact of the new $U(1)$ symmetry relevant to stability is on the running behavior of the top quark Yukawa coupling. In this paper we investigate  a new anomaly-free $U(1)^\prime$ for which the SM Higgs may carry charge.  Such a new symmetry may originate from  GUT models \cite{u13r} or string inspired models \cite{Anchirdiqyu}. Only right-handed fermions and the SM Higgs boson carry a $U(1)^\prime$ charge which we normalize to be multiples of `$a$'.  The even number of fermion doublets required by the global $SU(2)_L$ anomaly \cite{globalsu2} is provided by the SM.  The absence of axial-vector anomalies \cite{avector1,avector2,avector3} in the presence of the $U(1)^\prime$ and the gravitational-gauge anomaly \cite{anog1,anog2,anog3} requires that certain sums of the $U(1)^\prime$ charges vanish.  The right-handed fermions are assigned charges of $\pm a$ so these anomaly-free conditions are
\begin{eqnarray}
&& SU(3)_C^2 U(1)^\prime:  \hspace{1cm}  - 2 (a) - 2 (-a) =0 \; , \\
&& SU(2)_L^2 U(1)_Y: \hspace{1cm}  0\; ,\\
&& U(1)_Y^2 U(1)^\prime: \hspace{1cm }  -\left[  3 \left( {2\over 3}\right)^2 a  +3 \left( {1 \over 3} \right)^2  (-a)+ (-1)^2 (-a)\right] = 0 \; , \\
&& {U(1)^\prime}^2 U(1)_Y: \hspace{1cm} -a^2 \left[ 3 \times {2 \over 3} - 3 \times {1 \over 3} -1 \right]=0 \; , \\
&& U(1)^\prime: \hspace{2cm} -\left[a +(- a)\right] - 3 [ a+  (-a) ]=0 \; , \\
&& {U(1)^\prime}^3: \hspace{2cm} -[a^3 + (-a)^3] - 3 [a^3 + (-a)^3] =0 \; .
\end{eqnarray}
The $U(1)^\prime$ charge of the SM Higgs is fixed by the Yukawa interactions.  We list in table I the quantum numbers of the fields under the $U(1)^\prime$.  

\begin{table}[htbp]
\centering
\begin{tabular}{c|c|c|c|c|c|c|c |c}
\hline fields &$\ell  $ & $Q_L$ & $\nu_R$ & $E_R$ &$U_R$ & $D_R$& $H$ & $\phi$ \\
\hline$U(1)^\prime$ & $0$ & $0$ & $a$ & $-a$ & $a $ & $-a$ & $a$  & $Xa$\\
\hline 
\end{tabular}
\caption{ Quantum numbers of  fields under local $U(1)$.  }\label{aaa}
\end{table}

We have also included an additional scalar field $\phi$ which appears in table I.  The $U(1)^\prime$ can be spontaneously broken by the addition of this scalar $\phi$ with non-zero VEV and transforming as a singlet under the SM gauge group.  The number of generations of the scalar $\phi$, $n_\phi$, and its $U(1)^\prime$ charge -- written as a multiple $X$ of $a$ -- are not fixed by the requirement of anomaly cancellation or any interactions with the SM fields.  The impact of the resulting singlet-Higgs interactions on stability constitutes a special case of earlier work \cite{gondering,Cheung:2012nb,higgsportal,threshold}, so we do not consider it in detail here.\footnote{A detailed study of the electroweak precision measurement constraints on this model  is also beyond the scope of this paper}  We only focus on the impact of the $U(1)^\prime$ gauge interaction on the Higgs vacuum stability.  Note that the charge normalization $a$ can be absorbed by a redefinition of the new gauge coupling $g_4\rightarrow g_4/a$.  Taking the coefficient of the $H^\dag H\phi^\dag \phi$ operator to be negligible, the one-loop $\beta$-function of $\lambda$ can be written as 
\begin{eqnarray}
16 \pi^2 \beta_\lambda^{(1)} =16 \pi^2(\beta_\lambda^{(1)})_{\rm SM} + {3\over 4 } \left(  16 g_4^4 + {24 \over 5 } g_1^2 g_4^2 +8 g_2^2 g_4^2 \right) -12 g_4^2 \lambda \; ,
\end{eqnarray}
where the second and third terms are the contribution of the $U(1)^\prime$.  The $\beta$-function of $g_4$ can be written as 
\begin{equation}\label{eq:g4_beta}
16 \pi^2 \beta_{g^{}_4} = \left( {2 \over 3} \times 8 n_F + {2 \over 3} n_H + {1 \over 3} n_\phi  X^2\right) g^{ 3}_4 \equiv b_4 g_4^3  \; ,
\end{equation}
where $n_F$, $n_H$ and $n_\phi$ are the number of generations of fermions (3), the SM Higgs doublet (1), and additional $U(1)^\prime$-breaking singlet, respectively.  The new gauge interaction also affects the evolution of the Yukawa coupling of the top quark. Its $\beta$-function is given by 
\begin{eqnarray}
16 \pi^2 \beta_{t}^{(1)} =16 \pi^2 (\beta_{t}^{(1)})_ {{\rm SM}} - 3 g_4^{ 2} y_t \; .
\end{eqnarray}
The number of generations of scalars $n_\phi$, their $U(1)^\prime$ charge $X$, and the value of the coupling $g_4$ at the input scale are not totally arbitrary.  At one-loop order, it is straightforward to determine the scale at which $g_4$ has a Landau pole (we will leave a study of the two-loop effects of the new $U(1)^\prime$ symmetry for future work).  Solving Eq. (\ref{eq:g4_beta}) for $g_4(\mu)$ and equating the resulting denominator with zero, the Landau pole scale is found to be
\begin{equation}\label{eq:g4_landau_pole}
\Lambda_\mathrm{Landau} = \mu_0 \exp[16\pi^2/2b_4 g_4(\mu_0)^2]
\end{equation}
where $g_4(\mu_0)$ is the value of the gauge coupling at the input scale $\mu_0$.  By increasing $g_4(\mu_0)$, $n_\phi$, or $X$ (the latter two increase $b_4$), $\Lambda_\mathrm{Landau}$ decreases.  For certain choices it will be true that $\Lambda_{Landau} < M_{pl}$ and as $g_4\rightarrow\infty$ then surely $\lambda$ becomes non-perturbative.  Nonetheless, our numerical analysis shows that the choice of $g_4(\mu_0)$ has a more direct impact on the running of $\lambda$ than varying $n_\phi$ and $X$, so we fix $n_\phi=X=1$ to allow the greatest freedom in choosing the value of the $U(1)^\prime$ gauge coupling at the input scale while avoiding the Landau pole.

\begin{figure}[h!]
\includegraphics[width=11cm]{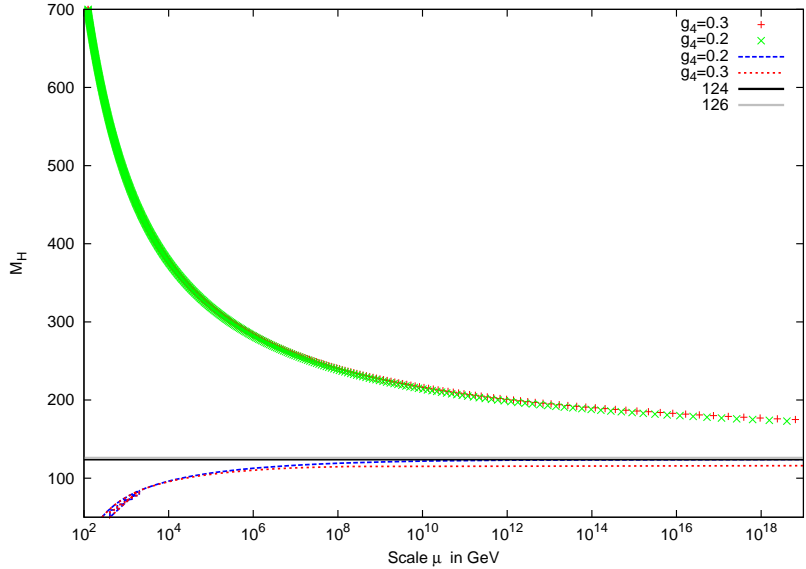}
\caption{ $M_H$ as the function of cutoff scale $\mu$. The plus and cross signs are the perturbativity constraint with $g_4=0.3, ~0.2$ respectively. The dashed and dotted lines are the vacuum stability constraint with $g_4 =0.2, ~0.3$ respectively. The horizontal band is the current measured value of Higgs boson mass.  }
\label{u1}
\end{figure}

We show in FIG. \ref{u1} the vacuum stability and perturbativity bounds on the SM Higgs mass $M_H$ as a function of $\mu$ for a set of values of $g_4$. We set the scale of $U(1)^\prime$-breaking at the $TeV$ scale and the mass of the corresponding $Z^\prime$ boson to be $M_{Z^\prime} \sim 2 ~{\rm TeV}$ in order to be roughly consistent with present LHC bounds\cite{zprime} . The plus and cross signs represent perturbativity constraints (by requiring $\lambda  < 8.2 $) with $g_4$ being $0.3$ and $0.2$,  respectively.  The dashed and dotted lines represent the Higgs vacuum  stability constraint with $g_4$ being $0.2$ and $0.3$, respectively. The horizontal band is the currently measured value of the Higgs mass.  It is evident from the figure that the SM Higgs vacuum stability predicts that   $g_4$ should be roughly larger than $0.2$ in this specific model, though a smaller value could also be viable if the impact of the singlet scalar-Higgs coupling is included. Alternatively, the presence of a sufficiently large gauge coupling would allow for stability in the absence of a significant singlet-Higgs interaction. In either case, such a parameter space might be accessible by the LHC, since we have assumed that $M_{Z^\prime} \sim 2~{\rm TeV}$. A detailed analysis of the collider signatures of the model will be presented in future work.  

\section{concluding remarks}

If the observed  $\sim 125$ GeV boson is, indeed, the Higgs boson, then stability of the SM electroweak vacuum up to the Planck scale may require that some BSM degrees of freedom become active at scales $\gsim$ 1 TeV. The presence of these degrees of freedom can modify the RG evolution of the Higgs quartic self-interaction that may otherwise run negative below the Planck scale. In general, three ingredients may change the running behavior of the Higgs self-coupling: four-scalar interactions, Yukawa interactions, and gauge interactions. 
In this paper, we have studied the SM Higgs vacuum stability problem in representative  minimal  extensions of the SM that also address other phenomenological problems,  accounting for the neutrino masses, dark matter from EW multiplets, and a new gauge symmetry.  We find that  vacuum stability of a $\sim$124-126~GeV Higgs up to the Planck scale could point to the existence of new scalars, which might be a TeV scale Higgs triplet in the type-II seesaw model, whose coupling $\lambda_4$ with the SM Higgs should be in the rather restricted range $(0.1,~0.4)$. Alternatively, stability could be achieved through existence of a series of  TeV-scale EW multiplets, which can be gauge-portal dark matter candidates, or to a new $U(1)^\prime$ gauge symmetry, in which  the SM Higgs carries non-zero charge.  In the former case, the number of new particles should be relatively large, {\em i.e.},  one generation quintuplet fermion (or  higher representational EW multiplet ) or at least two  quadruplet scalar and (or) triplet fermion multiplets.     If the new $U(1)$ symmetry is broken at the ${\rm TeV} $ scale,   then the SM Higgs vacuum stability implies that  $g_4$ cannot be arbitrarily small unless the couplings of the associated SM singlets to the Higgs doublet, the number of singlets, or the singlet $U(1)^\prime$ charge is sufficiently large.

\begin{acknowledgments}
This work was supported in part by DOE contracts DE-FG02-08ER41531 (MJRM and WCHAO) and DE-SC0007983 (MG), the Wisconsin Alumni Research Foundation (MJRM and WCHAO), and by the Wayne State University Division of Research (MG).

\end{acknowledgments}

\appendix

\section{$\beta$-functions of physics parameters }

The two loop $\beta$-function of the Higgs quartic coupling $\lambda$ is given by \cite{beta}
\begin{eqnarray}
{d \lambda \over d t } = {1 \over 16 \pi^2} \beta_\lambda^{(1)} + { 1 \over (16 \pi^2)^2} \beta_\lambda^{(2)} \; , 
\end{eqnarray} 
with
\begin{eqnarray}
\beta_\lambda^{(1)} &=&+12 \lambda^2 -\left(  {9 \over 5} g_1^2 + 9 g_2^2 \right) \lambda +  {27\over  100} g_1^4 + {9 \over 10} g_1^2 g_2^2 +  {9 \over 4 } g_2^4 + 12 \lambda y_t^2  -12 y_t^4 \; , \\
\beta_\lambda^{(2)} &=& -78 \lambda^3 + \left( {54 \over 5} g_1^2 + 54 g_2^2  \right) \lambda^2 - \left( {73 \over 8} g_2^4 -{117 \over 20}  g_1^2 g_2^2 - {1887 \over 200} g_1^4\right)\lambda -3 \lambda y_t^4 +{305 \over 8} g_2^6 
\nonumber\\ &&
- {289 \over 40} g_1^2 g_2^4 -{1677 \over 20 } g_1^4 g_2^2 -{3411 \over 1000} g_1^6 -64 g_3^2 y_t^4 -{16 \over 5} g_1^2 y_t^4 -{9 \over 2 } g_2^4 y_t^2 
\nonumber \\&&-
72 \lambda^2 y_t^2 + 10 \lambda  y_t^2 \left( {17 \over 20} g_1^2  + {9 \over 4 } g_2^2 + 8 g_3^2 \right) -  g_1^2 y_t^2\left( {171 \over 50 } g_1^2 -{63 \over 5 } g_2^2 \right)+ 60 y_t^6 \; .\label{lambdabeta}
\end{eqnarray}

The  two loop $\beta$-functions for the gauge coupling  are \cite{beta}:
\begin{eqnarray}
{ d g_i^{} \over d t  } = { b_i^{} \over 16 \pi^2 } g_i^3 + {1 \over (16 \pi^2)^2 } \left( \sum_{j=1}^3 b_{ij}^{} g_i^3 g_j^2 -\sum_{j= U,D,E} a_{ij}^{} g_i^3 {\rm Tr} [Y_j Y_j^\dagger ] \right)
\end{eqnarray}
with
\begin{eqnarray}
b_i^{}= \left( \matrix{ {41\over 10} & - {19 \over 6} & -7}\right)\; ,  \hspace{1cm }b_{ij} =\left( \matrix{{199 \over 50} &{27 \over 10} &~{44 \over 5} \cr {9 \over 10 } &{35 \over 6} &~12\cr {11 \over 10} &{9 \over 2 } & -26 }\right) \; ,  \hspace{1cm } a_{ij} =\left( \matrix{{17 \over 10} & {1 \over 2} & {3 \over 2} \cr {3 \over 2 } & {3 \over 2 } &{1 \over 2 } \cr 2 & 2 & 0} \right) \; .
\label{betag}
\end{eqnarray}

The two loop $\beta$-function of the top quark Yukawa coupling is \cite{beta}
\begin{eqnarray}
{d y_t \over d t} =  {y_t \over 16 \pi^2 } \beta_t^{(1)} + {y_t \over (16 \pi^2 )^2 } \beta_t^{(2)} \; , 
\end{eqnarray}
with
\begin{eqnarray}
\beta_t^{(1)} &=&+ { 9 \over 2} y_t^2 -\left( { 17 \over 20 } g_1^2 + {9 \over 4} g_2^2 + 8 g_3^2 \right) \; , 
\\ 
\beta_t^{(2)} &=& -12 y_t^4 +  y_t^2 \left( {393 \over 80} g_1^2 + { 225 \over 16 } g_2^2  + 36 g_3^2\right) + {1187 \over 600} g_1^4 - {9 \over 20 } g_1^2 g_2^2  + {19 \over 15} g_1^2 g_3^2 - {23 \over 4 } g_2^4
\nonumber \\ && 
+ 9 g_2^2 g_3^2 -108 g_3^4 + {3 \over 2 } \lambda^2 - 6 \lambda y_t^2 \; .
\label{betatop}
\end{eqnarray}

One loop $\beta$-functions of $\lambda_1$, $\lambda_2$ and $\lambda_5$ \cite{chao,schmidt}
\begin{eqnarray}
16\pi^2 \beta_{\lambda_1} = &&- \left(  {36\over 5} g_1^2 + 24 g_2^2 \right) \lambda_1 + {108 \over 25} g_1^4 + 18 g_2^4 + {72 \over 5} g_1^2 g_2^2  + 14 \lambda_1^2 + 4 \lambda_1\lambda_2  \nonumber \\&&+ 2\lambda_2^2 +4\lambda_4^2 + 4\lambda_5^2 +4 {\rm Tr}[Y_\Delta^\dagger Y_\Delta] \lambda_1 -8 {\rm Tr } [(Y_\Delta^\dagger Y_\Delta^{})^2] \; , \\
16\pi^2 \beta_{\lambda_2} = &&- \left(  {36\over 5} g_1^2 + 24 g_2^2 \right) \lambda_1 + 12 g_2^4 - {144 \over 5} g_1^2 g_2^2  + 3 \lambda_2^2 + 12 \lambda_1 \lambda_2 \nonumber \\&&-8\lambda_5^2 +4 {\rm Tr}[Y_\Delta^\dagger Y_\Delta] \lambda_1 +8 {\rm Tr } [(Y_\Delta^\dagger Y_\Delta^{})^2] \; , \\
 16\pi^2 \beta_{\lambda_5} = &&-\left({9\over 2} g_1^2  + {33\over 2} g_2^2 \right)\lambda_5 - {18 \over 5} g_1^2 g_2^2 + (2\lambda_1 -2\lambda_2 +2 \lambda+8\lambda_4 \nonumber \\ && +6y_t^2 +2 {\rm Tr } [Y_\Delta^\dagger Y_\Delta^{} ] ) \lambda_5^{} \; .
\end{eqnarray}


\begin{thebibliography}{99}


\bibitem{higgs}

F. Englert and R. Brout, Phys. Rev. Lett. {\bf 13}, 321 (1964); P. W. Higgs, Phys. Rev. Lett. {\bf 12}, 132 (1964); Phys. Rev. Lett. {\bf 13}, 508(1964); Phys. Rev.  {\bf 145}, 1156 (1966); G. S. Guralnik, Phys. Rev. Lett. {\bf 13}, 585(1964); T. W. B. Kibble, Phys. Rev. {\bf 155}, 1554 (1967).

\bibitem{atlas} 
  G.~Aad {\it et al.}  [ATLAS Collaboration],
  Phys.\ Lett.\ B
  [arXiv:1207.7214 [hep-ex]].
  
\bibitem{cms} 
  S.~Chatrchyan {\it et al.}  [CMS Collaboration],
  Phys.\ Lett.\ B
  [arXiv:1207.7235 [hep-ex]].

\bibitem{ewpm}

ALEPH, CDF, D0, DELPHI, L3, OPAL, SLD Collaborations, the LEP Electroweak Working Group, CERN PH-EP-2010-095, (2010). 


\bibitem{Holthausen:2011aa} 
  M.~Holthausen, K.~S.~Lim and M.~Lindner,
  JHEP {\bf 1202}, 037 (2012)
  [arXiv:1112.2415 [hep-ph]].


\bibitem{strumia}
  J.~Elias-Miro, J.~R.~Espinosa, G.~F.~Giudice, G.~Isidori, A.~Riotto and A.~Strumia,
  Phys.\ Lett.\ B {\bf 709}, 222 (2012)
  [arXiv:1112.3022 [hep-ph]].

\bibitem{xing2}
  Z.~-z.~Xing, H.~Zhang and S.~Zhou,
  Phys.\ Rev.\ D {\bf 86}, 013013 (2012)
  [arXiv:1112.3112 [hep-ph]].


  
\bibitem{Bezrukov:2012sa} 
  F.~Bezrukov, M.~Y.~.Kalmykov, B.~A.~Kniehl and M.~Shaposhnikov,
  arXiv:1205.2893 [hep-ph].
  
  
\bibitem{Alekhin:2012py} 
  S.~Alekhin, A.~Djouadi and S.~Moch,
  Phys.\ Lett.\ B {\bf 716}, 214 (2012)
  [arXiv:1207.0980 [hep-ph]].



\bibitem{Degrassi:2012ry} 
  G.~Degrassi, S.~Di Vita, J.~Elias-Miro, J.~R.~Espinosa, G.~F.~Giudice, G.~Isidori and A.~Strumia,
  JHEP {\bf 1208}, 098 (2012)
  [arXiv:1205.6497 [hep-ph]].

\bibitem{pdg}

K. Nakamura {\it et al}., (Particle Data Group), J. Phys. G {\bf 37},075021 (2010) and 2011 particle update for the 2012 edition.



\bibitem{wmap}

E. Komatsu, {\it et al}., arXiv:1001.4538[astro-ph.CO].

\bibitem{seesaw1}
P.~Minkowski,
  Phys.\ Lett.\ B {\bf 67}, 421 (1977);
  T.~Yanagida, in {\it Workshop on Unified Theories}, KEK report 79-18 p.95 (1979);
  M.~Gell-Mann, P.~Ramond, R.~Slansky,
  in {\it Supergravity} (North Holland, Amsterdam, 1979)
  eds. P.~van~Nieuwenhuizen, D.~Freedman, p.315;
  S.~L.~Glashow, in {\it 1979 Cargese Summer Institute on Quarks and Leptons} (Plenum Press,
  New York, 1980) eds. M.~Levy, J.-L.~Basdevant, D.~Speiser, J.~Weyers, R.~Gastmans and M.~Jacobs,
  p.687;
  R.~Barbieri, D.~V.~Nanopoulos, G.~Morchio and F.~Strocchi,
  Phys.\ Lett.\ B {\bf 90}, 91 (1980);
  R.~N.~Mohapatra and G.~Senjanovic,
  Phys.\ Rev.\ Lett.\  {\bf 44}, 912 (1980);
  G.~Lazarides, Q.~Shafi and C.~Wetterich,
  Nucl.\ Phys.\  B {\bf 181}, 287 (1981).

\bibitem{seesaw2}
W.~Konetschny and W.~Kummer,
  Phys.\ Lett.\  B {\bf 70}, 433 (1977);
%
 T.~P.~Cheng and L.~F.~Li,
  Phys.\ Rev.\  D {\bf 22}, 2860 (1980);
%
 G.~Lazarides, Q.~Shafi and C.~Wetterich,
 Nucl.\ Phys.\  B {\bf 181}, 287 (1981);
%
 J.~Schechter and J.~W.~F.~Valle,
  Phys.\ Rev.\  D {\bf 22}, 2227 (1980);
%
 R.~N.~Mohapatra and G.~Senjanovic,
  Phys.\ Rev.\  D {\bf 23}, 165 (1981).


\bibitem{seesaw3}
R.~Foot, H.~Lew, X.~G.~He and G.~C.~Joshi,
  Z.\ Phys.\  C {\bf 44}, 441 (1989).


\bibitem{Rodejohann2012} 
  W.~Rodejohann and H.~Zhang,
  JHEP {\bf 1206}, 022 (2012)
  [arXiv:1203.3825 [hep-ph]].
  
\bibitem{Masina:2012tz} 
  I.~Masina,
  arXiv:1209.0393 [hep-ph].

\bibitem{hexg}

X. G. He, G. C. Joshi, H. Lew and R. R. Volkas, Phys. Rev. D {\bf 44}, 2118 (1991); R. Foot, Mod. Phys. Lett. A {\bf 6}, 527 (1991); R. Foot, X. G. He, H. Lew and R. R. Volkas, Phys. Rev. D {\bf 50}, 4571(1994).


\bibitem{bminusl}

R. N. Mohapatra and R. E. Marshak, Phys. Rev. Lett. {\bf 44}, 1316 (1980).

\bibitem{Lindner:1988ww} 
  M.~Lindner, M.~Sher and H.~W.~Zaglauer,
  Phys.\ Lett.\ B {\bf 228}, 139 (1989).

\bibitem{hambye}

T. Hambye and K. Riesselmann, Phys. Rev. D {\bf 55}, 7255 ( 1996)

\bibitem{sirlin}

A. Sirlin and R. Zucchini, Nucl. Phys. B {\bf 266}, 389 (1986). 


\bibitem{beta}

M.E. Machacek and M. T. Vaughn, Nucl. Phys. B {\bf 222}, 83 (1983); Nucl. Phys. B {\bf 236}, 221 (1984); Nucl. Phys. B {\bf 249}, 70 (1985); C. Ford, I. Jack and D. Jones, Nucl. Phys. B {\bf 387}, 373 (1992); H. Arason {\it et al}., Phys. Rev. D {\bf 46}, 3945 (1992); V. Barger, M. S. Berger and P. Ohmann, Phys. Rev. D {\bf 47}, 1093 (1993); M. X. Luo and Y. Xiao, Phys. Rev. Lett {\bf 90}, 011601 (2003).


\bibitem{Barger:2007im} 
  V.~Barger, P.~Langacker, M.~McCaskey, M.~J.~Ramsey-Musolf and G.~Shaughnessy,
  Phys.\ Rev.\ D {\bf 77}, 035005 (2008)
  [arXiv:0706.4311 [hep-ph]].
  
\bibitem{Barger:2008jx} 
  V.~Barger, P.~Langacker, M.~McCaskey, M.~Ramsey-Musolf and G.~Shaughnessy,
  Phys.\ Rev.\ D {\bf 79}, 015018 (2009)
  [arXiv:0811.0393 [hep-ph]].

\bibitem{Cai:2011kb} 
  Y.~Cai, X.~-G.~He and B.~Ren,
  Phys.\ Rev.\ D {\bf 83}, 083524 (2011)
  [arXiv:1102.1522 [hep-ph]].


\bibitem{inertdm}
L. L. Honorez, E. Nezri, J. F. Oliver  and M. H. G. Tytgat, JCAP {\bf 0702}, 028 (2007)


\bibitem{raidal}
  M.~Kadastik, K.~Kannike, A.~Racioppi and M.~Raidal,
  JHEP {\bf 1205}, 061 (2012)
  [arXiv:1112.3647 [hep-ph]].


\bibitem{taiwan}
  W.~Chao,
  arXiv:1201.0364 [hep-ph];
  C.~-S.~Chen and Y.~Tang,
  JHEP {\bf 1204}, 019 (2012)
  [arXiv:1202.5717 [hep-ph]].



\bibitem{gondering}

M. Gonderinger, H. Lim and M. J. Ramsey-Musolf, Phys. Rev. {\bf D86} 043511 (2012); M. Gonderinger, Y. Li,  H. Patel and M. J. Ramsey-Musolf, JHEP 1001 002 (2010).  


\bibitem{Chun:2012jw} 
  E.~J.~Chun, H.~M.~Lee and P.~Sharma,
  arXiv:1209.1303 [hep-ph].
  
\bibitem{Cheung:2012nb} 
  C.~Cheung, M.~Papucci and K.~M.~Zurek,
  JHEP {\bf 1207}, 105 (2012)
  [arXiv:1203.5106 [hep-ph]].

\bibitem{higgsportal}
  O.~Lebedev,
  Eur.\ Phys.\ J.\ C {\bf 72}, 2058 (2012)
  [arXiv:1203.0156 [hep-ph]].

\bibitem{threshold}
  J.~Elias-Miro, J.~R.~Espinosa, G.~F.~Giudice, H.~M.~Lee and A.~Strumia,
  JHEP {\bf 1206}, 031 (2012)
  [arXiv:1203.0237 [hep-ph]].


\bibitem{chao}

Wei Chao and He Zhang, Phys. Rev. D {\bf 75}, 033003(2007).

\bibitem{schmidt}

M. A. Schmidt, Phys. Rev. D {\bf 76}, 073010 (2007).



\bibitem{Arhrib:2011uy}
  A. Arhrib, R. Benbrik, M. Chabab, G. Moultaka, M.C. Peyranere, et al., 
  Phys. Rev. {\bf D84} 095005 (2011).
  
  
\bibitem{perturbativity} 
  K.~Tobe and J.~D.~Wells,
  Phys.\ Rev.\ D {\bf 66}, 013010 (2002)
  [hep-ph/0204196].

\bibitem{Riesselmann:1996is}
  K. Riesselmann and S. Willenbrock, Phys. Rev. {\bf D55} 311 (1997).


\bibitem{minimal}

M. Cirelli, N. Fornengo and A. Strumia, Nucl. Phys. B {\bf 753}, 178 (2006). 

\bibitem{yicai}

Y. Cai, X. G. He and M. Ramsey-Musolf, JHEP {\bf 1112}, 054 (2011). 


\bibitem{Kumericki:2012bf} 
  K.~Kumericki, I.~Picek and B.~Radovcic,
  JHEP {\bf 1207}, 039 (2012)
  [arXiv:1204.6597 [hep-ph]].
  
\bibitem{FileviezPerez:2008bj} 
  P.~Fileviez Perez, H.~H.~Patel, M.~.J.~Ramsey-Musolf and K.~Wang,
  Phys.\ Rev.\ D {\bf 79}, 055024 (2009)
  [arXiv:0811.3957 [hep-ph]].


\bibitem{chaotri} 
  W.~Chao,
  arXiv:1202.6394 [hep-ph].


\bibitem{septuplet} 
  Y.~Cai, W.~Chao and S.~Yang,
  arXiv:1208.3949 [hep-ph].




\bibitem{u13r}
J. L. Hewett and T. G. Rizzo, Phys. Rept. {\bf 183}, 193(1989); P. Langacker, Phys. Rept. {\bf 72}, 185 (1981).

\bibitem{Anchirdiqyu} 
  L.~A.~Anchordoqui, I.~Antoniadis, H.~Goldberg, X.~Huang, D.~Lust, T.~R.~Taylor and B.~Vlcek,
  arXiv:1208.2821 [hep-ph];
  L.~A.~Anchordoqui, I.~Antoniadis, H.~Goldberg, X.~Huang, D.~Lust, T.~R.~Taylor and B.~Vlcek,
  arXiv:1206.2537 [hep-ph].


\bibitem{globalsu2}

E. Witten, Phys. Lett. B {\bf 177}, 324(1982)

\bibitem{avector1}

S. L. Adler, Phys. Rev. {\bf 177}, 2426(1969).

\bibitem{avector2}

J. S. Bell and R. Jackiw, Nuovo Cimento A {\bf 60}, 47(1969).

\bibitem{avector3}

W. A. Barden, Phys. Rev. {\bf 184}, 1848(1969).

\bibitem{anog1}

R. Delbourgo and A. Salam, Phys. Lett. B {\bf 40}, 381(1972).

\bibitem{anog2}

T. Eguchi and P. G. O. Freund, Phys. Rev. Lett {\bf 37}, 1251(1976).

\bibitem{anog3}

L. Alvarez-Gaume and E. Witten, Nucl. Phys. B {\bf 234}, 269(1984).


\bibitem{zprime} See ATLAS-CONF-2012-129.pdf and EXO-12-015-pas.pdf


\end{thebibliography}
\end{document}